\documentclass{cs23proc}

\usepackage{url}
\usepackage{hyperref}

\editors{Takeru Suzuki and the Cool Stars 23 Organizing Team}
\publisher{Zenodo}
\conference{The 23th Cambridge Workshop on Cool Stars, Stellar Systems, and the Sun (Cool Stars 23)}
\conferencedate{2026}

\title{Connecting Granulation and Magnetic Activity in Radial Velocities:\\The Next Breakthrough for High-Precision Spectroscopy}
\author{Ancy Anna John$^{1}$,
        Khaled Al Moulla$^{2}$,
        Federica Rescigno$^{1}$,
        Carmen San Nicolas Martinez$^{2}$,\\
        Andrew Collier Cameron$^{3}$,
        Thomas Wilson$^{4}$,
        Nadège Meunier$^{5}$,
        Sophia Sulis$^{6}$}

\affiliation{
$^{1}$ School of Physics \& Astronomy, University of Birmingham, Edgbaston, Birmingham B15 2TT, UK,\\
$^{2}$ Instituto de Astrofísica e Ciências do Espaço, Universidade do Porto, CAUP, Rua das Estrelas, 4150-762 Porto, Portugal\\
$^{3}$ Centre for Exoplanet Science, SUPA School of Physics and Astronomy,
University of St Andrews, North Haugh, St Andrews KY16 9SS, UK\\
$^{4}$ Department of Physics, University of Warwick, Gibbet Hill Road, Coventry CV4 7AL, UK\\
$^{5}$ Université Grenoble Alpes, CNRS, IPAG, F-38000 Grenoble, France\\
$^{6}$ Université Aix Marseille, CNRS, CNES, LAM, 13000 Marseille, France\\
}

\shorttitle{Connecting Granulation and Magnetic Activity in RVs}
\shortauthors{Anna John, Rescigno, Al Moulla \& San Nicolas Martinez}

\abs{Stellar variability has become the dominant limitation to achieving the radial velocity (RV) precision required for the detection and characterization of Earth-like exoplanets. While significant progress has been made in mitigating the effects of oscillations and magnetic activity, convective granulation and its interaction with stellar magnetic fields remain among the least understood sources of RV variability. To address these challenges, we organized the splinter session `Connecting Granulation and Magnetic Activity in Radial Velocities: The Next Breakthrough for High-Precision Spectroscopy' at Cool Stars 23. The session brought together researchers working on observations, numerical simulations, and data-driven techniques to discuss the current understanding of granulation-driven RV signals and identify the most promising directions for future progress. Through invited and contributed talks, followed by community discussions, participants emphasized the importance of combining physically motivated models with data-driven approaches, developing standardized benchmark datasets, and validating simulations against high-quality observations across a range of stellar types. The discussions also highlighted the need for coordinated observing strategies, improved characterization of individual spectral lines, and physically informed line-by-line analyses to disentangle convective and magnetic signals. This contribution summarizes the scientific discussions and community perspectives that emerged during the session and outlines the key challenges that must be addressed to reach the sub-40 cm s$^{-1}$ precision required for the next generation of RV planet searches.}

\begin{document}

\maketitle

\section{Introduction}

Variability phenomena on the surfaces of solar- and later-type stars produce signals in the measured radial velocity (RV). These signals evolve on timescales ranging from minutes, hours and days---induced by acoustic pulsations and convective granulation \citep[with disk-integrated amplitudes of $\sim$1\,ms$^{-1}$, e.g.,][]{Meunier+2015}---to months and years, associated with magnetically active regions and their growth patterns over stellar magnetic cycles \citep[amplitudes up to $\sim$10\,ms$^{-1}$, e.g.,][]{Meunier+2010}. To understand and model these signals to an accuracy within a few centimeters per second would be a remarkable feat in and of itself. However, in recent years, this challenge has additionally become a necessity in the field of exoplanetary sciences \citep{Crass+2021}. As pressure-stabilized, high-resolution spectrographs have become capable of measuring RVs with a precision of tens of centimeters per second \citep{Pepe+2014}, the main obstacle to detecting small-mass planets on long orbits is no longer instrumental. Instead, it is of stellar nature, as variability can drown out or mimic the periodic signals of potential companions. Earth-like planets, for instance, exhibit a semi-amplitude of only about $\sim$10${-}$40\,cms$^{-1}$. The inability to thoroughly account for stellar variability also impacts the accuracy of the inferred planetary properties and their subsequent atmospheric studies with, e.g., JWST.

While some activity components can be averaged out through an observational strategy \citep[such as pulsations;][]{Dumusque+2011,Chaplin+2019} or modeled quite accurately with the aid of simultaneously measured activity indicators \citep[such as active regions; see review by][]{Aigrain&Foreman-Mackey2023}, others have proven to be more difficult to account for. This includes granulation at several temporal and spatial scales \citep[cf. supergranulation at the larger end;][]{Rincon&Rieutord2018}, and especially its interplay with the evolving surface magnetic field structure. It has already been well established that magnetically active regions (i.e., faculae and spots) locally inhibit convective flows, causing asymmetries in the surface velocity field, which produces distinct RV signatures. Since these variations are quasi-periodic, the community is currently capable of isolating their contribution. Supergranulation on the other hand, due to its stochasticity at hard-to-probe timescales and its lack of known activity indicators, has been much more difficult to mitigate down to desired precision levels.

\section{Splinter Session Summary}

This splinter session had two primary goals:
\begin{enumerate}
    \item Summarize the recent developments in understanding the interplay between convective granulation and stellar surface magnetic fields.
    \item Identify the most urgent actions needed to improve our ability to isolate and de-correlate granulation effects in RV time series, enabling surveys of moderately active stars to achieve sub-40\,cms$^{-1}$ precision.
\end{enumerate}

We split the splinter session in two parts (divided by the coffee break), in which the first part was dedicated to cover the topic of stellar variability in RVs more broadly, and the second part covered how magnetic activity is coupled with granulation specifically. Each part consisted of an invited talk (15+5 mins), three contributed talks (12+3 mins), and a 25-minute discussion which included a short online poll using the Vevox platform\footnote{\url{https://vevox.app}}, where attendees could upload and vote on the most relevant questions to be discussed.

The talks and discussions are summarized below. For the talks, we only provide the names of the speakers and the titles of the talks; the PDFs of the presentation§s can be found on our splinter website\footnote{\url{https://eprvcoolstars23.github.io/program.html}}.

\subsection{Summary of Invited and Contributed Talks}

In the first half, we had the following talks:

\begin{itemize}
    \item Prof. Ignasi Ribas gave an invited talk, titled `The Star is the Signal: Taming Stellar Activity to Reveal Sun-Earth Analogues'.
    \item Dr. Jacob Luhn gave a contributed talk, titled `Resolving Oscillations and Granulation Within a Night with EPRV Spectrographs'.
    \item Dr. Manuel Perger gave a contributed talk, titled `Mitigating stellar RV jitter using orthogonal activity indices and a time-aware neural network'.
    \item Katlyn Hobbs gave a contributed talk, titled `Tracing the Spectral Fingerprints of Magnetic Activity Using Sun-as-a-star Observations'.
\end{itemize}

In the second half, we had the following talks:

\begin{itemize}
    \item Prof. Dainis Dravins gave an invited talk, titled `Radial-velocity microvariability: Sun and solar-type stars'.
    \item Varghese Reji gave a contributed talk, titled `Modelling the vertical velocity gradient to disentangle stellar activity from exoplanet signal'.
    \item Dr. Valeriy Vasilyev gave a contributed talk, titled `Sensitivity of spectral lines to granulation: from the Sun to K-type stars'.
    \item Dr. Ryan Rubenzahl gave the final contributed talk of the session, titled `Pixel-by-pixel response to granulation seen in EPRV solar spectra'.
\end{itemize}

\subsection{Summary of Discussions}

In the following subsections, we detail the discussed (most upvoted) questions and summarize the opinions, perspectives and consensus of the participants. For a list of all the submitted questions, see the Appendix.

\subsubsection{Discussion on Stellar Activity}

\begin{quote}
    \textit{Q1: Thanks to the latest techniques, and the higher amount and/or higher cadence of data (e.g. HARPS-N solar data, shown in Ignasi Ribas' talk), it seems we are starting to better understand stellar activity and ways to mitigate it. Additionally, and as mentioned during Jacob Luhn's talk, each star will need their own mitigation strategy. So my question is how should we handle future data observation strategy? If we need to observe a star multiple times a night, every nights, over multiple seasons, in order to fully understand and mitigate each of the different aspect of stellar noise, how should we decide which star to observe/study in this highly consuming observation method? Should we simply just stop observing stars all over the sky and only focus on a few of them, as a world- and instrument-wide strategy? But who decides which stars deserved to be observed?}
\end{quote}

The discussion emphasized the need for pre-survey characterization to ensure that limited high-precision RV resources are spent on the most promising targets. Since complementary observations such as multicolor photometry are relatively inexpensive, these should be used to identify and prioritize stars before committing to costly RV campaigns. Participants noted that the community will likely need to reach a consensus on which targets deserve intensive follow-up, for example by focusing on high-priority samples such as PLATO candidates. There was broad agreement that future RV surveys will need to be highly selective, while also building a broader understanding of how stellar activity manifests across different spectral types.

\begin{quote}
    \textit{Q2: Should the community prioritise better physical models or more data-driven / machine-learning approaches for activity mitigation?}
\end{quote}

The consensus was that both approaches are essential and complementary. Machine-learning and other data-driven methods can be powerful but must be carefully validated against observations to identify potential biases and degeneracies. Physical models provide important constraints that help prevent overfitting and improve interpretability. A particular concern is that purely data-driven approaches may inadvertently confuse planetary signals with stellar activity if the training data contain undiscovered planets. To mitigate this risk, participants highlighted the importance of training on synthetic datasets and using features that are, by construction, insensitive to planetary signals.

\begin{quote}
    \textit{Q3: Since most activity models are calibrated using the Sun, how well do these models generalize to active M-dwarfs, which are among the most important targets for habitable-zone planet searches?}
\end{quote}

The discussion suggested that the underlying physics governing stellar activity is expected to be similar across main-sequence stars, meaning that successfully reproducing solar observations increases confidence when extending models to M dwarfs. Participants noted that M dwarfs may, in some respects, even be simpler to model than solar-type stars, although important physical processes, such as supergranulation, are still not fully captured even for the Sun. Modern three-dimensional magnetohydrodynamic (MHD) simulations, including radiative transfer (e.g. MURaM), were highlighted as a promising approach because their complexity emerges naturally from first-principles physics. It was also noted that the larger radial velocity amplitudes of planets around M dwarfs partly compensate for the challenges posed by stellar activity.

\begin{quote}
    \textit{Q4: Should groups using solar data agree on a set of ``interesting'' time intervals to facilitate benchmarking between telescopes and mitigation methods?}
\end{quote}

There was strong support for developing standardized benchmark datasets to facilitate comparisons between different instruments and activity mitigation techniques. Such benchmarks should include observations spanning both high- and low-activity periods and would allow different methods to be tested under common conditions. Participants also highlighted the value of coordinated multi-site observing campaigns to achieve continuous, 24-hour solar coverage, enabling more robust benchmarking efforts.

\subsubsection{Discussion on Convective Granulation}

\begin{quote}
    \textit{Q1: Analyses of solar data seem to indicate that magnetic activity signals are easier to model than convective flows. Should we then be targeting more active stars for EPRV planet searches, since convection is inhibited by increased magnetic activity?}
\end{quote}

The discussion concluded that the answer is not straightforward. On the one hand, solar observations indicate that supergranulation evolves on shorter timescales during periods of higher magnetic activity, which could simplify aspects of activity modeling. On the other hand, increased magnetic activity introduces additional complications. Stronger magnetic fields produce larger starspots and more complex flow patterns, while changes in the stellar magnetic field can alter the star's physical structure over activity cycles. These effects may ultimately make activity more difficult, rather than easier, to predict and mitigate.

\begin{quote}
    \textit{Q2: How far can current MHD simulations realistically be trusted when extrapolating to other stars? One should not build science on `trust'. Take the model, compare with data. If they agree, probably it says something. If not, either model or observations are wrong.}
\end{quote}

Participants agreed that simulations should not simply be trusted but instead must be continually validated against observations. Rather than representing a true extrapolation, applying MHD models to other main-sequence stars is based on the expectation that the underlying physics evolves predictably with stellar parameters. Nevertheless, confidence in these models requires testing across a wide range of spectral types. Limitations of current simulations were also discussed, including incomplete treatment of fluid dynamics and magnetic turbulence, which affects the representation of small- and large-scale dynamos. Despite these shortcomings, current models reproduce many observable quantities remarkably well—particularly for the Sun—making them valuable tools even if they do not capture every aspect of the underlying physics. Participants also noted that full-star simulations remain computationally infeasible, requiring models to focus on representative regions instead.

\begin{quote}
    \textit{Q3: In the end, what will be the way to go (most promising directions) for solving our issue: looking into individual lines (special features, shape distortions rather than pure shifts), averaging over multiple lines (line specifically sensitive/insensitive to granulation/faculae), going into a line-by-line direction by using spectral line masks (physically informed line weights), and so on and so on... Any thoughts on this?}
\end{quote}

The consensus was that no single technique is likely to provide a complete solution. Instead, participants favored a combined strategy that incorporates multiple complementary approaches, including analyses of individual spectral lines, averaging over carefully selected line sets, and physically motivated line-by-line weighting schemes. Overall, the discussion suggested that the various methods currently being developed should be viewed as complementary rather than competing, with progress likely to come from integrating several techniques into a unified framework.

\section*{Acknowledgments}

The splinter organizers thank the Cool Stars SOC and LOC for their support and technical assistance. AAJ acknowledges funding from a UKRI Future Leader Fellowship, grant number MR/X033244/1. KA acknowledges support from the Swiss National Science Foundation (SNSF) under the Postdoc Mobility grant P500PT\_230225.
This article/publication is based
upon work from COST Action
CA22133, supported by COST
(European Cooperation in Science
and Technology).

\bibliographystyle{cs23proc}
\bibliography{bibliography.bib}

\begin{onecolumn}
\begin{appendix}

\section*{Appendix: Questions Uploaded by Attendees on Vevox Platform}

\vspace{0.5cm}

\subsection*{Discussion on Stellar Activity}

\noindent
Upvotes: 13\\
Discussed: Yes\\
Question:\\
\textit{Thanks to the latest techniques, and the higher amount and/or higher cadence of data (e.g. HARPS-N solar data, shown in Ignasi Ribas' talk), it seems we are starting to better understand stellar activity and ways to mitigate it. Additionally, and as mentioned during Jacob Luhn's talk, each star will need their own mitigation strategy. So my question is how should we handle future data observation strategy? If we need to observe a star multiple times a night, every nights, over multiple seasons, in order to fully understand and mitigate each of the different aspect of stellar noise, how should we decide which star to observe/study in this highly consuming observation method? Should we simply just stop observing stars all over the sky and only focus on a few of them, as a world- and instrument-wide strategy? But who decides which stars deserved to be observed?}\\

\noindent
Upvotes: 14\\
Discussed: Yes\\
Question:\\
\textit{Should the community prioritise better physical models or more data-driven / machine-learning approaches for activity mitigation?}\\

\noindent
Upvotes: 10\\
Discussed: Yes\\
Question:\\
\textit{Since most activity models are calibrated using the Sun, how well do these models generalize to active M-dwarfs, which are among the most important targets for habitable-zone planet searches?}\\

\noindent
Upvotes: 6\\
Discussed: Yes\\
Question:\\
\textit{Should groups using solar data agree on a set of ``interesting'' time intervals to facilitate benchmarking between telescopes and mitigation methods?}\\

\noindent
Upvotes: 7\\
Discussed: No\\
Question:\\
\textit{Besides RVs, what other types secondary or supplementary of data will be needed for correcting for stellar activity?}\\

\noindent
Upvotes: 6\\
Discussed: No\\
Question:\\
\textit{What is the best way to compare and benchmark techniques to address stellar variability? (e.g. rms, rate of detection, precision vs accuracy)}\\

\noindent
Upvotes: 10\\
Discussed: No\\
Question:\\
\textit{When using Gaussian Processes to remove stellar activity, how do you ensure that a low-amplitude planetary signal is not being partially absorbed by the activity model?}\\

\noindent
Upvotes: 6\\
Discussed: No\\
Question:\\
\textit{How do you see the future balance between physics-based models such as StarSim/MURaM and data-driven approaches like Gaussian Processes and machine learning? Can ML eventually replace physical modeling, or will it always need physical constraints?}\\

\noindent
Upvotes: 3\\
Discussed: No\\
Question:\\
\textit{Do we also need to focus on stellar activity characterization in multi-wavelength and multi-time to nearby Sun-like stars?}\\

\newpage

\noindent
Upvotes: 4\\
Discussed: No\\
Question:\\
\textit{Is there a common agreement on what will be the way to go in the end: LBL-RVs? GPs? Neural Networks? CCFs? And so on and so on ...}\\

\noindent
Upvotes: 3\\
Discussed: No\\
Question:\\
\textit{Whenever we try to manage stellar activity to such high precision, doesn't this also give us an opportunity to enter a new field of science of resolving stellar surface structures? Any thoughts on that ...}\\

\noindent
Upvotes: 2\\
Discussed: No\\
Question:\\
\textit{PARAS2 (2.5 m mount abu telescope) team in PRL, India is planning to upgrade the optical fibre of the instrument. ARIES, India is High Resolution Spectrograph (HRS) in 3.6 m Devasthal Opticl Telescope. If we can install a solar telescope in these instruments, we can observe 'sun as a star' for 24 hours}\\

\subsection*{Discussion on Convective Granulation}

\noindent
Upvotes: 6\\
Discussed: Yes\\
Question:\\
\textit{Analyses of solar data seem to indicate that magnetic activity signals are easier to model than convective flows. Should we then be targeting more active stars for EPRV planet searches, since convection is inhibited by increased magnetic activity?}\\

\noindent
Upvotes: 4\\
Discussed: Yes\\
Question:\\
\textit{How far can current MHD simulations realistically be trusted when extrapolating to other stars? One should not build science on `trust'. Take the model, compare with data. If they agree, probably it says something. If not, either model or observations are wrong.}\\

\noindent
Upvotes: 7\\
Discussed: Yes\\
Question:\\
\textit{In the end, what will be the way to go (most promising directions) for solving our issue: looking into individual lines (special features, shape distortions rather than pure shifts), averaging over multiple lines (line specifically sensitive/insensitive to granulation/faculae), going into a line-by-line direction by using spectral line masks (physically informed line weights), and so on and so on... Any thoughts on this?}\\

\noindent
Upvotes: 2\\
Discussed: No\\
Question:\\
\textit{How well GP works for granulation: is the granulation signal property consistent with the asteroseismic measurements? Are there some differences?}\\

\noindent
Upvotes: 3\\
Discussed: No\\
Question:\\
\textit{What is the role and need of co-ordinated observations to understand super/granulation. What have we learned so far, what are the existing gaps?}\\

\noindent
Upvotes: 3\\
Discussed: No\\
Question:\\
\textit{How to observationally test spectral line response to granulation or activity from atomic parameters, ionisation state etc.}\\

\end{appendix}
\end{onecolumn}

\end{document}